\documentclass[aps,prl,twocolumn,showpacs,floatfix,amssymb,amsmath]{revtex4}
\usepackage{graphicx}
\usepackage{dcolumn}
\usepackage{bm}
\usepackage{color}
\newcommand{\be}{\begin{equation}}
\newcommand{\ee}{\end{equation}}
\newcommand{\bea}{\begin{eqnarray}}
\newcommand{\eea}{\end{eqnarray}}
\newcommand{\p}{\partial}

\newcommand{\vare}{\varepsilon}

\newcommand{\re}{\mbox{e}}
\newcommand{\ba}{\begin{array}}
\newcommand{\ea}{\end{array}}
\def\nn{\nonumber\\}
\def\vare{{\varepsilon}}
\def\up{\uparrow}

\begin{document}
  
  \title{Unscreened Coulomb interactions and quantum spin Hall phase in neutral zigzag graphene ribbons}
  
  \author{Mahdi Zarea } 
  \author{Carlos B\"{u}sser} 
  \author{Nancy Sandler}
  \affiliation{Department of Physics and Astronomy, Nanoscale and Quantum Phenomena Institute, and Condensed Matter and Surface Science Program,\\Ohio University, Athens, Ohio 45701-2979}
  
  \date{\today}
  
  \begin{abstract}

A study of the effect of unscreened Coulomb interactions on the quantum spin Hall (QSH) phase of finite-width neutral zigzag graphene ribbons is presented. By solving a tight-binding Hamiltonian that includes the intrinsic spin-orbit interaction (I-SO), exact expressions for band-structures and edge-states wavefunctions are obtained. These analytic results, supported by tight-binding calculations, show that chiral spin-filtered edge states are composed of localized and damped oscillatory wavefunctions, reminiscent of the ones obtained in armchair ribbons. The addition of long-range Coulomb interactions opens a gap in the charge sector with a gapless spin sector. In contrast to armchair terminations, the charge-gap vanishes exponentially with the ribbon width and its amplitude and decay-length are strongly dependent on the I-SO coupling. Comparison with reported ab-initio calculations are presented. 

  \end{abstract}
 
  \pacs{81.05.Uw,73.20.At, 73.43.-f, 85.75.-d}   
  \maketitle


Graphene, the single-layer carbon material isolated for the first time in 2004\cite{Gaim,Kim} provides a unique window to explore phenomena usually associated with QED as well as quantum Hall physics\cite{physicstoday,Neto1}. Its peculiar band-structure, with linear spectra at specific points in momentum space, allows for a description of low-energy properties in terms of a Dirac Hamiltonian for massless electrons. Besides the interest on understanding its fundamental properties, graphene has received increasing attention due to its potential technological applications. In particular, much effort is focused on obtaining graphene nanoribbons (GNR) with tailored properties\cite{Kim2,Avouris,Tapaszto} that could be key elements in future electronics and spin-transport applications\cite{spin-transport}. 

Ribbons present an exceptional opportunity to study confinement effects and enhanced Coulomb interactions on graphene properties. Moreover, they are a natural test-bench to investigate the physics of the quantum spin Hall phase (QSH) originated by intrinsic-spin orbit (I-SO) interactions \cite{Kanemele}. This phase has been studied in the low-energy limit (Dirac Hamiltonian) and was found to be characterized by a bulk insulating behavior and chiral spin-filtered edge states. Tight-binding and perturbative renormalization group calculations \cite{Kanemele, Moore,Onari}, predict the phase to be stable against disorder and screened Coulomb interactions along the edges. These arguments also suggest a transition to an insulating phase for unscreened interactions. 
However, at present no detailed study exists on its properties and the fate 
of the spin-filtered edge states when the  {\it full} graphene 
band-structure is considered and {\it inter-edge} 
 unscreened Coulomb interactions are included.  

In this letter we present the first study of the QSH phase in zigzag GNR (ZGNR) that includes the combined effects of finite-system size and electron-electron interactions. The analytic solution of a tight-binding Hamiltonian with I-SO interactions provides exact expressions for the band-structure and wavefunctions that characterize the QSH phase giving insight into the nature of the spin-filtered edge states. To investigate the effects of unscreened Coulomb interactions we exploit the linear nature of the spectrum and use bosonization methods to analyze the Hamiltonian.  Our results indicate that a gap opens in the charge sector while the spin sector remains gapless. Within the harmonic approximation, we provide an expression for the gap in terms of the parameters of the model. The analysis of the data suggests an exponential vanishing gap with increasing ribbon's width, with an amplitude and decay-length dependent on the I-SO coupling constant. These findings are in contrast with results for armchair ribbons\cite{Mehdi} and provide an interpretation for reported DFT calculations\cite{Louie2}.
\begin{figure}[!]
\includegraphics[width=.5\textwidth]{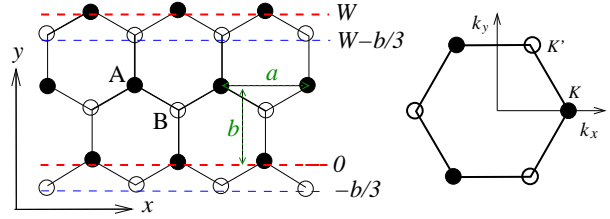}
\caption{A zigzag graphene nanoribbon (ZGNR) (left) of width $W$ and its corresponding Brillouin zone (right). Dotted lines indicate positions where boundary conditions are applied. }
\label{gra1}
\end{figure}

\texttt{The Hamiltonian:}   
In the tight-binding description, the Hamiltonian $H_0=\sum_{<ij>}[t{c}_{i}^{\dag}{c}_j+h.c]$ ($i,j$ label position and spin degrees of freedom); is written in a spin-dependent basis, the \emph{pseudo-spin basis} $\Psi^{\dagger, s}=\left( u^s_A; u^s_B \right)$, where the left (right) component represents the amplitude of the wavefunction at a lattice $A (B)$ site, and $s={\pm}$ labels the spin. The intrinsic spin-orbit interaction involves spin-dependent second-neighbor hopping terms: 
$H_{I-SO} \sim i t' (\vec{d_{ik}} \times \vec{d_{kj}})c^{\dag}_is^zc_j$, 
with $\vec{d_{ik}}$ as first neighbor lattice vectors connecting electrons at positions $i$ and $k$, $c^{\dagger}_{i}$ the electron creation operator at site $i$ and ${s^z}$ the spin\cite{Kanemele,Mehdi}. This term satisfies all the symmetries of the graphene lattice\cite{Manes} and plays a crucial role in magnetic properties of finite-width ribbons. Reported values for $t'$ from ab-initio calculations range from $1.2K$ to $10mK$\cite{str}.
 In reciprocal space the Hamiltonian reads:
\be
H  =  \left( \begin{array}{cc}
  s\gamma & \phi  \\
  \bar\phi & -s\gamma  \end{array}\right)
\label{eq:hamiltonian}
\ee
where $\gamma(k_x,k_y) =  2t'(\sin k_xa -2\sin{\frac{k_xa}{2}}\cos{k_yb})$, $\phi(k_x,k_y) =  t( \re^{ik_y2b/3}+2\cos{\frac{k_xa}{2}}\re^{-ik_yb/3})$, $\bar\phi(k_x,k_y)=\phi(k_x,-k_y)$, and $b=a\sqrt{3}/2$. The corresponding eigenvalues and eigenstates are $E=\pm\vare=\pm\sqrt{\phi\bar\phi+\gamma^2}$ and:
\bea
\Psi_{\pm}^{s}&=&N\left( \begin{array}{c}
  1  \\
  \eta_{\pm}^{s}\end{array}\right)\re^{ik_yy}\re^{ik_xx}
\label{eq:energy-bulk}
\eea
where $\eta_{\pm}^{s}(k_x,k_y)=(\pm\vare - s\gamma)/\phi=\bar\phi/(\pm\vare + s\gamma)$. $N$ is the normalization factor. A gap opens at the Dirac points and the system becomes a bulk-insulator\cite{Kanemele}. 

To study confined geometries we consider a ribbon of length $L$ along the $x-$direction, with finite width $W$ along $y$ (see Fig.(\ref{gra1})), and impose hard-wall boundary conditions on the wavefunctions of Eq.(\ref{eq:hamiltonian}). In addition to the set of boundary conditions used in zigzag ribbons: ($\varphi_A(0)=\varphi_B(W-b/3)=0$)\cite{Fujita,Nakada,Brey,Hikihara}, the I-SO interaction introduces new ones: $\varphi_A(W)=\varphi_B(-b/3)=0$. The more general wavefunction is a superposition of four degenerate states of the form of Eq.(\ref{eq:energy-bulk}), with $k_y=(\pm k_1; \pm k_2)$. To guarantee their degeneracy, the parameters must satisfy 
the condition: $\cos k_1b+\cos k_2b=2c(1-\tau^2)$ with $\tau^2=1/(8t'^2\sin^2 {\frac{k_xa}{2}})$. A wavefunction for a right mover with spin up is given by:
\bea
\Psi_{+\up}&=N&\re^{ik_xx}\Bigg[a\left( \begin{array}{c}
    1  \\
    \eta_{1, \up}  \end{array}\right)\re^{ik_1y}
  +b\left( \begin{array}{c}
    1  \\
    {\bar\eta_{1, \up}}  \end{array}\right)\re^{-ik_1y}\nn
  &&+c\left( \begin{array}{c}
    1  \\
    {\eta_{2, \up}}  \end{array}\right)\re^{ik_2y}
  +d\left( \begin{array}{c}
    1  \\
    {\bar\eta_{2, \up}}  \end{array}\right)\re^{-ik_2y}
  \Bigg]
\label{main-wf}
\eea
with $\eta_{j,\up}=\eta(k_y=k_{j})_{\up}$ and $N$ is the normalization factor. The new dispersion relation and the coefficients appearing in Eq.(\ref{main-wf}) are obtained from the condition:
\bea
\big((\gamma_1-\gamma_2)^2+(\phi'_1-\phi'_2)(\bar\phi'_1-\bar\phi'_2)\big)
\sin^2(k_{-}W/2)&&   \nn
=\big((\gamma_1-\gamma_2)^2+(\phi'_1-\bar\phi'_2)(\bar\phi'_1-\phi'_2)\big)
\sin^2(k_{+}W/2) &&
\label{det-c}
\eea
where $\phi'_{j}=\phi_{j}\re^{i k_{j}b/3}$, $\gamma_{j}=\gamma(k_{y}=k_{j})$, and $k_{\pm}=(k_{1} \pm k_{2})/2$. The degeneracy condition plus Eq.(\ref{det-c}) uniquely determine them as functions of $t,t'$ and $W$. 
\begin{figure}[!]
  \includegraphics[width=.5\textwidth]{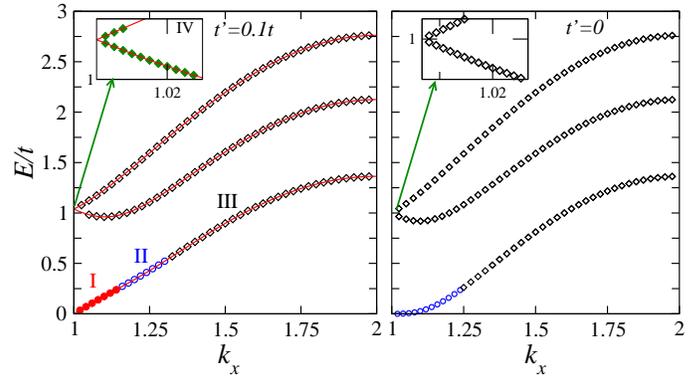}
  \caption{Band structure of ZGR with $W=4b$ in the presence (left panel) and absence (right panel) of I-SO interactions with $t'=0.1t$. Top inset shows the change in the bulk-bands. Full lines are numerical tight-binding calculations.}
  \label{gra}
\end{figure}
Fig.(\ref{gra}) shows the band-structure of a ribbon of $W=4b\approx 1 nm$ with $t'=0.1t$, in the range $\pi \leq k_x\leq 2\pi$. The quasi-degeneracy of the lower band is lifted for $\pi \leq k_{x} \leq 4\pi/3$ by the I-SO interaction as predicted \cite{Fujita,Nakada,Hikihara,Brey}. The spectrum remains gapless at $k_{x}=\pi$, with linear dispersions and right- and left-mover states for each spin component \cite{Kanemele}. The nature of the edge states depends on $k_x$: in region $I$ (near $k_{x}=\pi$), the wavenumbers $k_{1/2}$ are complex, rendering damped oscillatory wavefunctions; 
in region $II$ (extending to $\Delta k_{x} \sim 1/W$ before the Dirac
points) however, the wavefunction becomes fully localized at the edges. The predicted 'spin-filtered edge states' are formed by a combination of states from these two regions. As $k_{x}$ is increased further, $k_{2}\to 0$ while $k_{1}$ is purely imaginary, resulting in oscillatory wavefunctions in region $III$. A similar analysis reveals that bulk conduction bands are made from a combination of localized and oscillatory wavefunctions, except near $k_x=\pi$ where wavefunctions are fully localized (see inset in Fig.(\ref{gra})). Notice that a simplified expression for the lower-band dispersion can be found when $t'/t\gg b/W$: $\vare=\pm6t'\sin(k_xa)\sqrt{1+16t'^2\sin^2(k_xa/2)}$ which near $k_x=\pi$ can be approximated by $\vare\approx6t'k_xa$ with the velocity of right (left) movers given by
$\hbar v=\pm 6t'a$. These results hold for ribbon widths $W=2Mb$ with $M$ integer. For $W=(2M+1)b$ the hard-wall boundary condition breaks one of the sublattices discrete translation invariance symmetry along the $y-$direction, which opens a gap $\Delta \approx t(2t'/t)^{W-1}$ at $k_x=\pi$ and $t'/t\ll1$. 

\begin{figure}[!]
  \includegraphics[width=.5\textwidth]{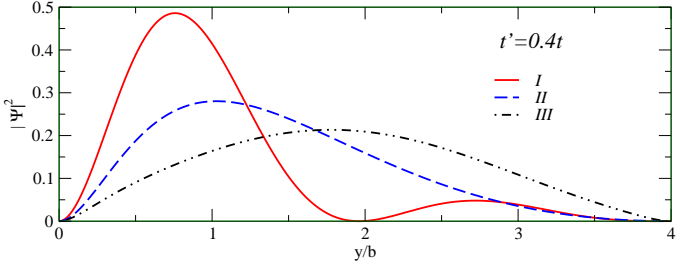}
  \caption{Probability distribution for a spin-up right-mover for values of $k_{x}$ in region I: $\pi \leq k_{x}a \ll \pi +t'/t$;  II: $\pi + t'/t \ll k_{x}a \ll 4\pi /3 - b/W$, and III: $k_{x}a \gg 4\pi /3 - b/W$. Curves for spin-down right-movers are obtained reflecting through  the $y=W/2$ plane ($y/b \to (W-y)/b, W=4b$).}
  \label{prb}
\end{figure}

The coefficients of the wavefunction (Eq.(\ref{main-wf})) are given by $a(k_1,k_2)=f^{1/4}\re^{-ik_1W/2}$ where
\bea
f&=&\frac{(\bar\eta'_1-\eta'_2)(\bar\eta'_1-\bar\eta'_2)
  (\vare+\gamma_1)(\sin k_2b\sin k_2W)}
{(\eta'_1-\eta'_2)(\eta'_1-\bar\eta'_2)
  (\vare+\gamma_2)(\sin k_1b\sin k_1W)}
\eea
with $\eta'_j=\eta_j\re^{-ik_jb/3}$, $b=a(-k_1,k_2), c=a(k_2,k_1)$ 
and $d=a(-k_2,k_1)$. Representing the resulting spin-up right-mover with
$\Psi^{\dag}_{+\up}=\Psi^{\dag}_{R\up}=\re^{-ik_xx}(\varphi_A(y),\varphi_B(W-y))$
 the corresponding spin-up left-mover wavefunction takes the form 
$\Psi^{\dag}_{-\up}=\Psi^{\dag}_{L\up}=\re^{-ik_xx}( -\varphi_B(W-y),\varphi_A(W-y))$ and those for spin-down right/left-movers are obtained by taking $t'\to -t'$. Fig.(\ref{prb}) shows the probability distribution for a spin-up right (spin-down left) mover as a function of the position $y$ across the ribbon for different values of $k_x$.  The predicted localized\cite{Kanemele} wavefunction along the edge changes from exponential to damped oscillatory as $k_{x}$ moves towards $k_x =\pi$. This oscillatory behavior for $k_{x} \simeq \pi$ is reminiscent of the one obtained in armchair ribbons (AGNR) with I-SO interactions suggesting an effect independent of particular edge termination \cite{Mehdi}. Also, notice that there is no spin-accumulation along the edges neither net magnetization in equilibrium.

{\bf Coulomb interactions:}    
For neutral isolated graphene ribbons, screening of the Coulomb potential is suppressed due to low electron densities. Starting from the un-screened Coulomb potential $U_{pp'}(\Delta x,\Delta y)={e^2}/{\kappa\sqrt{a_0^2+\Delta x^2+\Delta y_{pp'}^2}} $ with $p=+$ ($p=-$) labeling sub-lattice $A$ ($B$) and $\Delta y=y-y'+\delta_{p,-p'}a/\sqrt{3}$ \cite{Gogolin}; the interacting Hamiltonian is reduced to
\be
H=\frac{1}{4} \sum_{s,s'} \int d^{2}x d^{2}y U^{\pm}
{\bm \rho}^{\dag}_{\pm,s}(x,y){\bm \rho}_{\pm,s'}(x',y')
\ee
where ${\bm \rho}_{\pm,s}={\bm\rho}_{As}\pm{\bm\rho}_{Bs}$ with ${\bm \rho}_{ps}={\bm c}^{\dag}_{ps}{\bm c}_{ps}$ and  $U^{\pm}=U_{AA}\pm U_{AB}$. 
 The linear-dispersion of the lower-band near $k_{x} \simeq \pi$, makes possible to use the bosonization technique to study intra-band Coulomb scattering on spin-filtered edge states.  
Near $k_x=\pi$, the lower (edge) bands are separated by
($\Delta E\sim t \sim 2.4 eV$) from the upper bands. This gap reduces near the Dirac points to 
$\Delta E \sim 1/W$ but does not vanish and, at low enough temperatures, inter-band transitions can be neglected.
We introduce a spin-dependent Bogoliubov transformation 
from the original site-operators to a right/left-movers basis:
\bea
{\bm c}_{A\up}(k_x,y)=\varphi_A(k_x,y){\psi}_{R\up }
-\varphi_B(k_x, W-y){\psi}_{L\up}&&\nn
{\bm c}_{B\up}(k_x,y)=\varphi_B(k_x,y){\psi}_{R\up}
+ \varphi_A(k_x, W-y){\psi}_{L\up}&&
\eea	  
where ${\psi}_{R/L\up}={\psi}_{R/L\up}(k_x)$ (for 
spin-down: $\varphi_{A/B}(y)$ is replaced by $\varphi_{B/A}(W-y)$). Unlike previous works \cite{Wu, Moore} this model includes  scattering processes {\it between} the edges represented by a marginal operator term in the Hamiltonian. As a consequence, the non-local irrelevant terms are neglected
and the resulting Hamiltonian is:
\bea
&&{\cal H}=\frac{\hbar v_{c/s}}{2}
\sum_{c,s}\big[\frac{1}{K_{c/s}}(\p_x\Phi_{c/{s}})^2
  +K_{c/s}(\p_x\Theta_{c/{s}})^2\big] \nn
&&+\frac{m}{\pi a}[\cos\sqrt{8\pi}\Phi_{s}-\cos(\sqrt{8\pi}\Phi_c+4k_Fx)]
\label{Hboson}
\eea
where $K_{c/s}=\sqrt{(1+Q_{c/s}/\hbar v)/(1+R_{c/s}/\hbar v)}$; $v_{c/s}=v\sqrt{(1+R_{c/s}/\hbar v)(1+Q_{c/s}/\hbar v)}$, and $k_F$ is the Fermi momentum. The parameters in these expressions are:
\bea
R_s&=&\int \frac{d^2y}{2\pi}\sum_{u=\pm} U_0^u(\Delta y)B^u(y)B^u(y')\nn
R_c&=&\int \frac{d^2y}{2\pi}\sum_{u=\pm} U_0^u(\Delta y)F_+^u(y)F_+^u(y')-R_s\nn
Q_s&=&\int \frac{d^2y}{2\pi}\sum_{u=\pm} U_0^u(\Delta y)F_-^u(y)F_-^u(y')
\label{parameters}
\eea
with $Q_c=0$, $m=R_s/2a$, $\Delta y = (y-y')$ and
\bea
&&F^{u}_{\nu}(y)=\frac{1}{2}[\varphi_A(y)\varphi_A(y)+u\varphi_B(y)\varphi_B(y)\nn
&&+\nu\varphi_B(W-y)\varphi_B(W-y)+u\nu\varphi_A(W-y)\varphi_A(W-y)]\nn
&&B^{u}(y)=\varphi_A(y)\varphi_B(W-y)-u\varphi_B(y)\varphi_A(W-y)
\label{FB}
\eea
In the expressions above, $U_0^u$ (zero Fourier component) is the unscreened part of the Coulomb interaction.  Eq.(\ref{Hboson}) contains forward scattering terms ($F^{u}_{\nu}$) that preserve chirality 
and modify charge and spin-velocities $v_{c/s}$ and Luttinger parameters $K_{c/s}$.
Backward scattering terms ($B^{u}$) originate from the mixing of same spin right- and left-movers located at opposite edges and produce the mass term $m$. Using standard approximations \cite{Bbook}, we obtain:
\be
\Delta=\frac{\hbar v_c}{a}\left(\frac{2a m}{ \hbar v_c}\right)^{1/(2-2K_c)}.
\ee
\begin{figure}[t]
  \includegraphics[width=.5\textwidth]{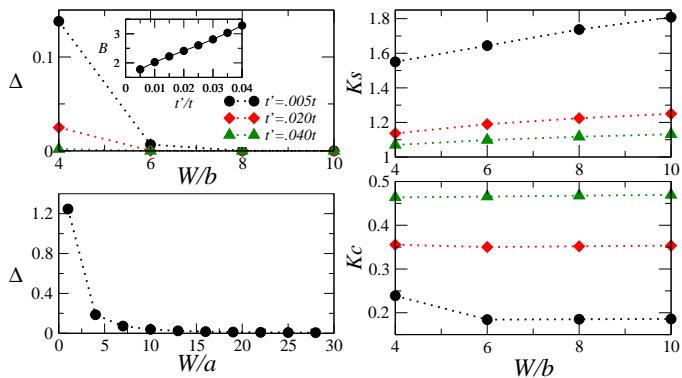}
  \caption{Left pannels: Charge gap $\Delta$ (in units of eV) as a function of ribbon width $W$ for a zigzag ribbon of length $L \simeq 25 \mu m$ (upper). The armchair (lower) ribbon gap is plotted for comparison (length $L \simeq 400 \mu m$). Inset in ZGNR pannel: exponent $B$ (in units of $b$) defined by $\Delta\sim A\re^{-BW}$ as a function of $t'/t$. Right pannels: Luttinger parameters $K_s$ (up) $K_c$(down) as function of ribbon width.}
  \label{gap}
\end{figure}
Fig(\ref{gap}) shows the dependence of the gap as a function of $W$ for different values of $t'$ as well as the Luttinger parameters $K_{c/s}$ indicators of charge and magnetic orders. The data was obtained taking $\kappa=2.45$ (appropriate for graphene over a $SiO_{2}$ substrate) and $a_0\approx a$. 
An analysis of the width-dependence reveals an exponentially vanishing gap $\Delta\sim A\re^{-BW}$ (see inset in Fig.(\ref{gap})) with gap amplitude $A$ and decay-length $1/B$ decreasing with increasing $t'$.  These results can be understood in terms of the renormalization of the Coulomb interaction, as measured by the dimensionless parameter $g=e^2/\kappa \hbar v$, with $v$ determined by $t'$. For $t'=0$ our results agree with previous work\cite{Hikihara}. As $t'$ is increased, same spin right- and left-movers wavefunction overlaps are suppressed. The reduced overlap diminishes the back-scattering terms and decreases the gap. Thus, larger values of $t'$ render
{\it gapless} chiral-spin filtered states even in the presence of electron-electron interactions. In this regime, the system is described by models with no inter-edge Coulomb interactions\cite{Wu, Moore}. The data also show: $K_c<1$ and $K_s>1$ for all values of $t'$ considered, indicating a strong tendency to form magnetically ordered structures at the edges. This conclusion is in agreement with ab-initio calculations that revealed strong magnetic instabilities (of ferro or anti-ferromagnetic nature) for ZGNR \cite{magnetic-instability}.
As Fig.(\ref{gap}) shows, a comparison with AGNR reveals gaps of an order of magnitude smaller in zigzag ribbons. Furthermore, 
the gap dependence $\Delta \sim 1/W$ (see Fig.(\ref{gap})) for AGNR changes to $\Delta \sim e^{-W}$; and reflects the different nature of the respective edge states. For AGNR, the edge states are surface states (not localized along the edges) formed by a linear superposition of extended states near Dirac points\cite{Brey}. The I-SO interaction mildly localizes them along the edges with a localization length {\it inversely} proportional to the I-SO coupling \cite{Mehdi, Sengupta} and makes them spin-filtered states. In contrast, the edge states in ZGNR are strongly localized {\it in the absence} of $t'$ and the main effect of the I-SO interaction is to make them spin-filtered states with a suppressed unscreened Coulomb repulsion.   

In summary, we have obtained exact expressions for band-structure and wavefunctions of a ZGNR in its QSH phase, by solving a tight-binding Hamiltonian that includes intrinsic-spin orbit interactions. For appropriate ribbon widths, the I-SO interaction induces spin-filtered edge states with linear dispersion around $k_x=\pi$. These states are composed by localized and damped-oscillatory wavefunctions and closely resemble those found in AGNR. Unscreened Coulomb interactions open a charge-gap in the spectrum while the spin-sector remains gapless as in AGNRs. However, numerical results suggest an exponentially vanishing gap as a function of the ribbon's width $W$ with gap amplitudes and decay-lengths decreasing with increasing values of the I-SO coupling $t'$. The exponential dependence of the gap is the first one reported for finite-width zigzag ribbons. Our estimates give smaller values for gap amplitudes than those predicted in DFT calculations\cite{Louie2}, while in general agreement with the trends observed in them. Furthermore, they emphasize the relevance of inter-edge unscreened Coulomb interactions in the QSH phase, as compared with previous results based on renormalization group arguments. The results also suggest an enhanced magnetic order along the edges reminiscent to the one found in AGNR for large values of the I-SO coupling \cite{Mehdi}. These results have immediate experimental consequences and could be tested in  current setups \cite{Tapaszto}. 

This work was supported by the Ohio University Postdoctoral Research Fellowship (MZ) and NSF-DMR 0710581 (NS). We acknowledge support from the Aspen Center for Physics where this work was written up.

\end{document}